\documentclass[useAMS,usenatbib]{mn2e}
\usepackage {graphicx}
\usepackage {times}
\usepackage{pdflscape}
\usepackage{rotating}
\usepackage{natbib}

\title[No Ly$\alpha$ emitters around a QSO at z=6.4
]{No Ly$\alpha$ emitters detected around a QSO at z=6.4: Suppressed by the QSO?\footnotemark[0]
}
\author[Goto]{Tomotsugu Goto$^{1}$ 
,
 Yousuke Utsumi$^{2}$,
Satoshi Kikuta$^{3}$,
Satoshi Miyazaki$^{3,4}$,
Kensei Shiki$^{5}$,
\newauthor
Tetsuya Hashimoto$^{1}$
\\
$^{1}$National Tsing hua University, No. 101, Section 2, Kuang-Fu Road, Hsinchu, Taiwan 30013\\
$^{2}$Hiroshima Astrophysical Science Center, Hiroshima University, 1-3-1 Kagamiyama, Higashi-Hiroshima, Hiroshima, 739-8526, Japan\\
$^{3}$
Department of Astronomical Science, SOKENDAI (The Graduate University
for Advanced Studies), Osawa, Mitaka, Tokyo 181-8588, Japan\\
$^{4}$National Astronomical Observatory, 2-21-1 Osawa, Mitaka, Tokyo
181-8588,Japan\\
$^5$Department of Physical Science, Hiroshima University, Higashi-Hiroshima, Hiroshima 739-8526, Japan\\
}

\begin{document}
\def\Hg{H$\gamma$}
\def\Hd{H$\delta$}

\date{\today; in original form 2011 August 9}

\pagerange{\pageref{firstpage}--\pageref{lastpage}} \pubyear{2009}

\maketitle

\label{firstpage}

 \begin{abstract}
Understanding how QSO's UV radiation affects galaxy formation is vital to our understanding of reionization era.
Using a custom made narrow-band filter, $NB906$, on Subaru/Suprime-Cam,
 we investigated the number density of Ly$\alpha$ emitters (LAE) around a QSO at z=6.4.
 To date, this is the highest redshift narrow-band observation, where LAEs around a luminous QSO are investigated. Due to the large field-of-view of Suprime-Cam, our survey area is $\sim$5400~cMpc$^2$, much larger than previously studies at z=5.7 ($\sim$200 cMpc$^2$).

  In this field, we previously found a factor of 7 overdensity of Lyman break galaxies (LBGs). Based on this, we expected to detect $\sim$100 LAEs down to $NB906$=25 ABmag. 
  However, our 6.4 hour exposure found none.
  The obtained upper limit on the number density of LAEs is more than an order lower than the blank fields.
  Furthermore, this lower density of LAEs spans a large scale of 10 $p$Mpc across. 
  A simple argument suggests a strong UV radiation from the QSO can suppress star-formation in halos with $M_{vir}<10^{10}M_{\odot}$ within a $p$Mpc from the QSO, but the deficit at the edge of the field (5 $p$Mpc) remains to be explained.
 \end{abstract}


\section{Introduction}

 Quasars (QSOs) are expected to be a tracer of high-density regions because they harbor supermassive black holes with masses up to $\sim 10^{10} M_{\odot}$, which are likely to reside in massive dark matter halos of $\sim 10^{13} M_{\odot}$. Such massive hales are theoretically expected to hold many massive galaxies, and evolve into present-day massive clusters although with a significant scatter \citep[e.g.,][]{2005Natur.435..629S,RO09}.

 Observationally, at $2<z<5$, there have been both positive  \citep[e.g.,][]{2003ApJ...596...67D,2012MNRAS.422.2980S,2015MNRAS.452.2388H} and negative detection \citep[e.g.,][]{2004MNRAS.353..301F,K07} of overdensities around luminous QSOs.
  Recent discoveries of increasing number of $z\sim6$ QSOs advanced the research into higher-z, again to find mixed results.
  Some reported discoveries of overdensities around QSOs \citep{2005ApJ...622L...1S,2006ApJ...640..574Z}, while others failed to find any significant overdensity \citep{2009ApJ...695..809K,B13,2017ApJ...834...83M}.
  See \citet{2016A&ARv..24...14O} for a more complete review.
  


 Why do results vary?
 A part of the reason could be a small field of view (FoV) of the previous instruments, especially HST/ACS, which most previous work relied on at $z\sim6$.
 To rectify the situation, in our previous work, we took advantage of a large FoV of the Subaru Suprime-Cam to find a factor of 7 overdensity of Lyman break galaxies (LBGs) around a luminous QSO at z=6.4, with a ring-like distribution centred on the QSO with a radius of $\sim$3 Mpc \citep{U10}.
Such a large scale structure could not be found without the large FoV of the Suprime-Cam.

 However, the result was based on small statistics of 7 objects. Photometric redshifts of LBGs have a large uncertainty. Spectroscopic confirmation would be ideal, but the LBGs are faint ($z_R<25.1$ and $z'<25.4$). Spectroscopic confirmation is not trivial, especially when their Ly$\alpha$ emission is weak.
 
 An efficient alternative is to use a narrow-band filter and detect Ly$\alpha$ emitters (LAEs). Being an imaging observation, one can survey a large area at once, focusing on the Ly$\alpha$ emission line, where we expect a high S/N ratio. 

 For this purpose, we have developed a custom-made narrow-band filter, $NB906$, to detect LAEs at z=6.4.
 This is the highest redshift, where the narrow-band imaging technique is used to investigate environment of a QSO with a large field of view of Supreme-Cam.
 The QSO-galaxy interaction can be more clearly seen at higher-z, where galaxies have less chance to form before the QSO turns on.

 In this paper, we report our $NB906$ observation of the QSO field centred on CFHQS J2329$-$0301\citep[Table \ref{tab:targets};][]{2007AJ....134.2435W} at z=6.417 \citep{2010AJ....140..546W}. Unless otherwise stated, we adopt the following cosmology: $(h,\Omega_m,\Omega_L) = (0.7,0.3,0.7)$.

\begin{table*}
 \begin{minipage}{180mm}
  \caption{Target information adopted from \citet{2007AJ....134.2435W,Goto09,Goto11}. }\label{tab:targets}
  \begin{tabular}{@{}lcccccccc@{}}
  \hline
   Object &  z$_{\rm Mg II}$&              $i'_{AB}$ & $z'_{AB}$ & $z_{R}$ & $NB906$ & $J$                                  &  \\ 
 \hline
 \hline
   CFHQS J232908.28-030158.8  & 6.417$\pm$0.002 & 25.54$\pm$0.02 & 21.165$\pm$0.003&  21.683$\pm$0.007   &    20.20$\pm$0.003       & 21.56$\pm$0.25 \\
\hline
\end{tabular}
\end{minipage}
\end{table*}

%
%


\section{Observation}
\label{Observation}

We have developed a custom-made narrow-band filter $NB906$, whose central wavelength is 905nm (z = 6.4 for Ly$\alpha$) for Subaru/Suprime-Cam's F/1.86 convergence light, with FWHM of 15.8nm ($\Delta$z = 0.1).  

Using this $NB906$ filter, we observed a field centred on QSO CFHQS J2329$-$0301 at z=6.4 with Subaru Suprime-Cam. In this field, we have previously obtained deep images in $i',z',$ and $z_R$ filters \citep{U10}. 
The $z_R$ filter covers the redder side of the $z'$ band and has a central wavelength of 9900\AA \citep{shimasaku05}.
The total exposure time of the $NB906$ observation was 23044  sec.
After masking the halos and horizontal spikes surrounding bright stars, as well as the outer edge of the image, the effective area is 0.219 deg$^2$. 
The data reduction was performed using the {\ttfamily imcat}\footnote{https://www.ifa.hawaii.edu/$\sim$kaiser/imcat/}.
The FWHM of the combined image was 0.64''. 
Exposure times of all filters are summarized in Table \ref{tab:exptime}.


\begin{table}
\begin{center}
\caption{
Summary of data used.
}
\label{tab:exptime}
\begin{tabular}{lcll}
  \hline
 Band &  PSF size (arcsec) & Exp time (s) & Limiting mag\\
 & & & (3 $\sigma$AB, 1.2'' aperture)\\
 \hline
   \hline
 $NB906$ & 0.64 & 23044 & 25.73\\ 
$i'$ &   0.61 & 3600  & 26.95\\
$z'$ &   0.60 & 6900 & 26.13\\
$z_R$  & 0.62 & 12532 & 25.46\\
 \hline 
\end{tabular}
\end{center}
\end{table}

\section{Analysis and Results}\label{analysis}

\subsection{LAE selection}


We measure magnitudes using MAG\_AUTO of SExtractor and
$1.2''$-diameter aperture. The aperture size is chosen to be twice as large as the seeing FWHM to increase the S/N ratio. 
We adopt MAG\_AUTO as total magnitudes, while 
we use a $1.2''$-diameter aperture magnitude 
to measure colours of objects in order to obtain
colours of faint objects with a good signal-to-noise ratio.
We correct the magnitudes of objects for Galactic extinction of $E(B-V)=0.0384$ \citep{SFD98}.

\begin{figure}
\begin{center}
  \includegraphics[scale=0.55]{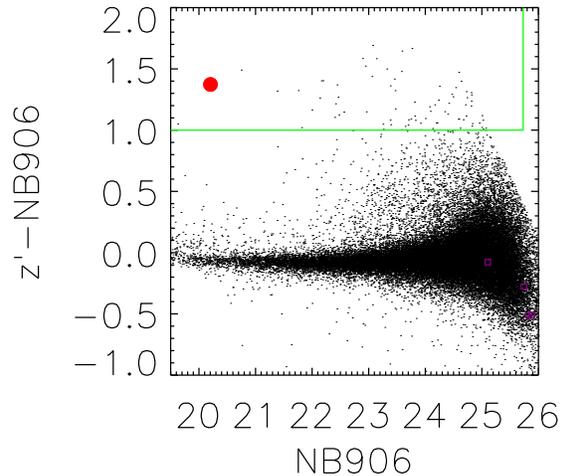}
\end{center}
\caption{
 The aperture $z'-NB906$ colour against $NB906$ best magnitude ({\tt MAG\_AUTO}).
 The red point is the colour of CFHQS J2329$-$0301. The black dots are all objects with {\ttfamily flag=0}. The purple squares are LBGs identified in \citet{U10} (not all of them were detected in $NB906$). The green lines show our selection criteria. Note that there exist objects with $z'-NB906>1.0$ but not selected as LAEs because of bluer $i'-z'$ colour.
}\label{fig_cmd}
\end{figure}

\begin{figure}
\begin{center}
   \includegraphics[scale=0.55]{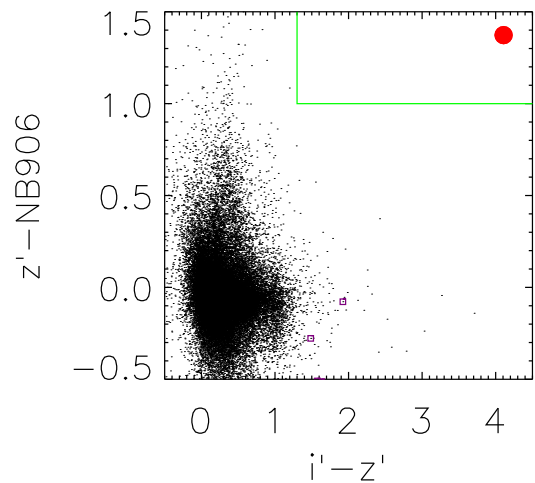}
\end{center}
\caption{
 $i'-z'-NB906$ colour-colour diagram.
 The red point is the colour of CFHQS J2329$-$0301. The black dots are all objects with {\ttfamily flag=0}. The purple squares are LBGs identified in  \citet{U10}.
 The green lines show our selection criteria.
}\label{fig_iznb}
\end{figure}

To reduce contamination by false detection, we only use the ``detected cleanly'' and ``no blending'' objects that have {\ttfamily FLAGS=0} in SExtractor. 
We plot all such objects with the black dots in Figure \ref{fig_cmd},
where the $z'-NB906$ colour is shown as a function of  $NB906$ magnitude.
Figure \ref{fig_iznb} presents a $i'-z'-NB906$ colour-colour diagram based on the $NB906$-detection catalog.

Following previous work \citep{T05,O10}, 
we select LAEs with the narrow-band excess, 
and the existence of lyman break as follows,
{\footnotesize
\begin{eqnarray}
\label{eq:selectionNB906}
z'-NB906>1.0, \ i'-z'>1.3,\  {\rm and}  
 ~NB906<NB906_{\rm lim,3\sigma}.
\end{eqnarray}
}
When objects are not detected in the $z'$ band, we used 3 $\sigma$ limiting magnitude to evaluate the objects.

\begin{figure}
\begin{center}
  \includegraphics[scale=0.5]{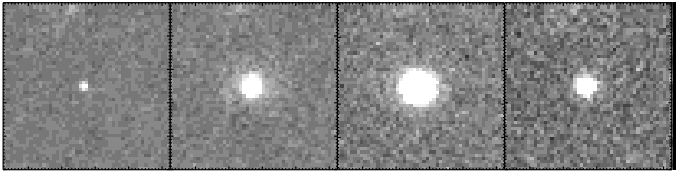}
\end{center}
\caption{ 10$\times$10'' images of CFHQS J2329$-$0301 at z=6.4. Filters are $i',z', NB906$, and $z_R$ from left to right.}
 \label{fig:qso}
\end{figure}

 With these criteria, we did not find any LAE across the FoV of the Suprime-Cam,
 except the QSO CFHQS J2329-0309, which is shown with the red circle in Figs.\ref{fig_cmd}-\ref{fig_iznb}.
Fig.\ref{fig:qso} shows 10$\times$10'' images of the QSO.
 There were 6 $NB906$ detection without $i'$ or $z'$-band detection. We have eye-balled them to find none of them were real objects (either a cosmic-ray, or affected by the edge of the field).

Note that FWHM of the $NB906$ filter is 15.8nm, which is larger than that of $NB927$ (13.2nm) used in \citet{T05,O10}. With the same colour criteria of 1.0, our selection criteria correspond to LAEs with the rest-frame equivalent width, EW$_0$, of 43\AA, instead of 36\AA~of \citet{T05,O10}.
 In an attempt to match the criteria in terms of  EW$_0$, we tried $z'-NB906>0.8$, which selects LAEs with EW$_0>$36\AA. Yet no LAE was found (see \citet{Schenker14} for an example EW distributions of LAEs).

\begin{figure}
\begin{center}
  \includegraphics[scale=0.55]{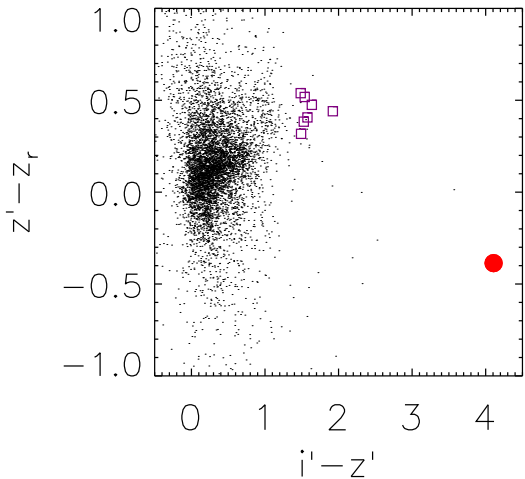}
\end{center}
 \caption{
 The red point is the colour of CFHQS J2329$-$0301. The black dots are all objects with {\ttfamily flag=0}. The purple squares are LBGs identified in  \citet{U10}.
   Our  LAE selection criteria are  $i'-z'>1.3$, $z'-NB906>1.0$, and $NB906<26.0$.
}\label{fig_izzr}
\end{figure}

\citet{U10} detected 7 LBGs within the field.
All of them are identified in this work, with the same criteria of $i'-z'>$1.3, and $z'-z_R>$0.3, as shown with purple squares in Fig. \ref{fig_izzr}. 
These LBGs are also shown in Figs.\ref{fig_cmd}-\ref{fig_iznb}.
Fig. \ref{fig_iznb} shows none of them has significant excess in $NB906$, suggesting their EWs are less than 30\AA (This assumes that they have redshifts with Ly$\alpha$ falling in the NB filter).

\subsection{Completeness}
\begin{figure}
\begin{center}
 \includegraphics[scale=0.55]{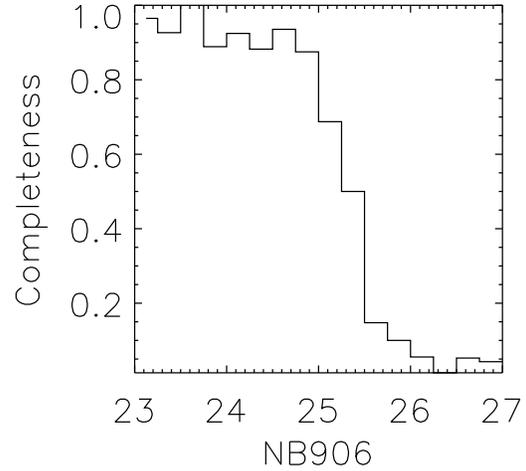}
\end{center}
\caption{
Completeness as a function of NB906 magnitude.
}\label{fig:compl}
\end{figure}

 We estimate the detection completeness of $NB906$
 image as a function of narrow-band magnitude.
 We construct a composite PSF using stars in the NB906 image.
 Then, we randomly distribute 1000 artificial PSFs with varying magnitude between 23.0 and 27.0 onto the NB906 image. We did not remove regions affected by bright stars. Therefore, the measured completeness includes the effective area correction, which is expected to be a few percent.
 Next, we try to detect them in the same manner as the detection of our LAEs with SExtractor. 
We plot the detection completeness as a function of
 $NB906$ magnitude in Figure \ref{fig:compl}.
The detection completeness drops to $\simeq 50$\% at around $NB906$=25.4 mag.


\subsection{LAE luminosity function}

\begin{figure}
\begin{center}
     \includegraphics[scale=0.55]{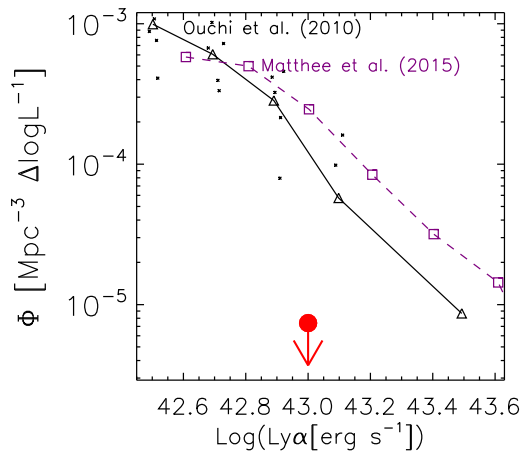}
\end{center}
 \caption{ LAE LF.  One $\sigma$ upper limit from our work with $NB906$ is shown with the red arrow. Note the bin size is 1 dex.
 The black and purple lines show results from \citet{O10} and  \citet{2015MNRAS.451..400M} at z=6.6.
 The black small dots are results from five different sub-fields in \citet{O10}.
 }\label{fig:LF}
\end{figure}

Next, we compare the number density of LAEs with those in the blank fields.
We obtain the upper limit of the number density of LAE by simply dividing the number counts of LAEs by the effective survey comoving volume, defined as the FWHM of the bandpass (15.8 nm) times the area of the survey. 
Resulting one $\sigma$ upper limit is shown in Fig.\ref{fig:LF}. For a comparison, LAE LF at z=6.6 is shown with the black triangles \citep{O10}. Our upper limit is lower by about an order than the LAE density in the fields.


\section{Discussion}\label{discussion}

\subsection{Notes on LBGs}

We also identified 7 LBGs in \citet{U10}, and plotted their colours in Figs.\ref{fig_cmd}-\ref{fig_iznb} with the purple squares. However, none of them has $z'-NB906>1$ to satisfy the criteria to be LAEs. Four of them are not even detected in $NB906$. This means either they are not in the redshift range of $NB906$ ($6.4<z<6.5$), or they are but without a strong Ly$\alpha$ emission. Because their broad-band colours are still very red ($i'-z'>1.9$, and $z'-z_R>0.3$), they are likely to be at $z>5.8$.
Previous observations reported the fraction of bright LBGs with a strong Ly$\alpha$ emission is small at z$>6$.
For example, \citet{2011ApJ...728L...2S,2011ApJ...743..132P,Schenker14} reported the fraction of bright LBGs (EW$>$25\AA, M$_{UV}<$-20.25) with a Ly$\alpha$ emission is $\sim$20\% at z$\sim$6, and $\sim$10\% at z$\sim$7. Our LBGs are even brighter with  M$_{UV}$=$-$22.2$\sim-$21.7. The fraction would be even smaller. It is not too surprising if none of the six LBGs had a strong Ly$\alpha$ emission.
 Further conclusions need spectroscopic confirmation of these LBGs.

%

  \subsection{Comparison to previous work}

  In previous work investigating the environment of QSOs, there exist both positive and negative results on the detection of overdensity of galaxies around QSOs. At lower redshift, QSOs' duty cycles become relatively shorter compared with the age of the Universe. There are increasing chances that surrounding galaxies may have formed before the central QSO did. In addition, LBGs and LAEs are different in their mass, and thus, physical effects from the central QSOs might be also different \citep[e.g.,][]{K07}.
  Therefore, to simplify the discussion, below we compare with results at higher-z (z$>$4), which used LAEs to investigate the QSO environment.

In previous work,
\citet{K07} found that LBGs without Ly$\alpha$ emission form a filamentary structure near QSO SDSS J0211$-$0009 at z=4.87, while Ly$\alpha$ emitters are distributed around\footnote{Note that \citet{2011ApJ...739...56D} used MgII line to measure the redshift of J0210$-$0009 to be z=4.894. If so, Ly$\alpha$ emissions could be out of the $NB711$ filter.}
but avoid QSO within a distance of $\sim$770 $p$kpc.
 \citet{2012MNRAS.422.2980S} used the Taurus Tunable Filter to find a
significant galaxy overdensity around a QSO at z=4.528 over 35 arcmin$^2$.
\citet{2017ApJ...834...83M} and \citet{B13} investigated the environment of two z=5.7 quasars to find no enhancement of LAEs in comparison with blank fields.
\citet{kikuta} observed environments of LAEs around two QSOs at z$\sim$4.9 using a narrow-band filter. They found that  two QSOs are located near local density peaks ($<2\sigma$), but the number densities of LAEs in a larger spatial scale  are not significantly different from those in blank fields.



 While results of the previous studies vary, we would like to highlight differences in our work.
 Our QSO is at the highest redshift of z=6.4. At a higher redshift, the age of the Universe is younger.  With less time available for halo formation, the effect of the QSO UV radiation would be clearer. 
 This could be a reason why  we found the number density of LAEs is even smaller than the general field by 3 $\sigma$.  It is an important future task to investigate QSO luminosity dependence of LAE distribution using multiple QSOs at similar or higher redshifts.

 Also previous work at z$\sim$5.7 could not rule out the QSO environment could be overdense at large scale ($\sim$10 $p$Mpc) because of their smaller field of view ($\sim$200 cMpc$^2$ at most).
Our larger area coverage of  $\sim$5400 cMpc$^2$, for the first time at z$\sim$6, ruled this out by finding  the lower density of LAEs is over a large scale of $\sim$10 $p$Mpc across.

\subsection{Physical interpretation}

By finding the lack of LBGs near the QSO, \citet{U10} discussed the strong UV radiation may have suppressed the formation of galaxies in the vicinity of the QSO.
 The QSO is associated with a giant Ly$\alpha$ nebulae \citep{Goto09,Goto11}, reflecting a strong UV radiation from the QSO.
 In this work, we found the lack of LAE not just in the QSO vicinity but in the whole field.
 
 \citet{K07} quantitatively estimated how much QSO's UV radiation can suppress such galaxy formation.
 QSO CFHQS J2329$-$0301's absolute magnitude is $M_{1450\AA}$=-25.2 \citep{2007AJ....134.2435W}, which is about 1.2 magnitude fainter that in \citet{K07} at z=4.87 ($M_{1450\AA}$=-26.4).
 Following their arguments, within 770 $p$kpc, $J_{21}\sim24$. The QSO can suppress star-formation (SF) in halos with $M_{vir}<10^{10}M_{\odot}$, while halos with  $M_{vir}>10^{11}M_{\odot}$ are almost unaffected (see their Fig.8). 
 However, at the edge of the field ($\sim$5$p$Mpc away), $J_{21}$ is $\sim0.6$.  Only SF in halos with  $M_{vir}<10^{9}M_{\odot}$ can be suppressed. 
  Previous estimates of halo mass of typical LAEs ($L_{{\rm Ly \alpha}}\sim10^{42} {\rm erg/s}$) are around $M_{vir}\sim10^{10}M_{\odot}$ \citep{2007ApJ...671..278G,O10,2015MNRAS.450.1279G}.
  If so, the SF can be suppressed in halos near the QSOs, but it remains to be explained why LAEs are not detected around the edge of the field, where UV radiation is weaker.

   There are several notable sources of uncertainty on the discussion. On the theoretical side, it was assumed that stars form at the centre of a spherical halo. If star formation takes place after a disk-like collapse or in substructures, the impact of photoionization will be greater and the inferred mass of the host halo can be larger. For example, some evidence was found that high-z sub-millimeter galaxies are rotationally-supported \citep[e.g.,][]{Goto_CO}.
      
  On the observational side, we should note that halo mass estimates of LAEs depends on the age of the stellar population, and thus there remains uncertainty.   
  If the halo mass of LAEs are $M_{vir}<10^{9}M_{\odot}$, the lack of LAEs in our QSO field is consistent with the suppression scenario.
 
  The suppression scenario could explain the detection of LBGs in \citet{U10}. LBGs are thought to have older stellar population, and thus more massive than LAEs \citep{2006ApJ...648L...5O}.
The host halo masses of typical LBGs are inferred to be one order of magnitude larger than those of LAEs from angular auto-correlation function measurements \citep{2007ApJ...671..278G,2015MNRAS.450.1279G}.
  The mass estimate of our LBGs is also uncertain because we do not have deep near-infrared data. However considering bright magnitude, they are likely to be massive galaxies with  $M_{vir}=10^{11-12}M_{\odot}$.
  If so, they could survive the UV radiation from the QSO
  (However, also see \citet{2012MNRAS.421.2543B}, who argue more massive halos up to 1.2$\times 10^{12}M_{\odot}$ can be suppressed).
   Also being older galaxies, LBGs could have formed before the QSO turned on, while LAEs could not form after the QSO formation.

  We should, however, keep in mind other possibilities.
  QSO's radiation is thought to be strongly beamed. More suppression is expected in the beaming directions, while we do not see any angular dependence in LAEs and LBGs distributions.
  Among known z$>$6 QSOs, our QSO is relatively faint, with $M_{1450\AA}$=-25.2 and a black hole mass of 2.5$\times$10$^8 M_{\odot}$ \citep{2010AJ....140..546W}.
  The host galaxy of the QSO was not detected by ALMA \citep{2013ApJ...770...13W}, putting tight constraints on the SFR in the QSO host. It may therefore be possible that the QSO itself is in a low mass halo and therefore does not reside in a high density peak in the early Universe \citep{RO09}.
  

    Another possibility is that these LAEs in the field are dusty, and not detected in $NB906$. It has been known that high-z ULIRGs and sub-millimeter galaxies are dusty and faint in optical, despite their large star-formation rate \citep[e.g.,][]{Goto15CFHT}. If such galaxies are in the field, they could have escaped our observation.

    However, to conclude any further, we need more reliable data in both quality and quantity. This work presented the first case at $z>6$, where a QSO environment was investigated in a scale of $\sim$10 Mpc with a narrow-band filter. Even including studies with smaller FoV, there are only a few more examples of the narrow-band studies at z$\sim$6, due to the rarity of QSOs and few available windows for narrow-band observations. With the emergence of recent large surveys, much larger number of high-z QSOs are being found \citep{100QSO}, some of which have redshift corresponding to the narrow-band filter's \citep{B13}. For example, recently discovered are several QSOs at z=6.6 \citep{TG}, whose Ly$\alpha$ emission can be observed with a narrow-band filter, $NB927$. 
 It will be important future work to investigate environment of such QSOs using the narrow-band technique to obtain statistically robust results. 



\section*{Acknowledgments}

We thank the anonymous referee and R. Overzier for many insightful comments.
We are very grateful to all of the Subaru Telescope staff.
We acknowledge grant aid for the narrow band filter from the Department of Astronomical Sciences of the Graduate University for Advanced Studies (SOKENDAI).
YU was supported by JSPS KAKENHI Grant Number JP26800103, JP24103003.
 TG acknowledges the support by the Ministry of Science and Technology of Taiwan through grant NSC 103-2112-M-007-002-MY3, and 105-2112-M-007-003-MY3.

\bibliography{tomo_qso_v3} 

\begin{thebibliography}{37}
\expandafter\ifx\csname natexlab\endcsname\relax\def\natexlab#1{#1}\fi

\bibitem[{Ba{\~n}ados} et~al.(2013){Ba{\~n}ados}, {Venemans}, {Walter}, {Kurk},
  {Overzier} \& {Ouchi}]{B13}
{Ba{\~n}ados} E., {Venemans} B., {Walter} F., {Kurk} J., {Overzier} R., {Ouchi}
  M., 2013, \apj, 773, 178

\bibitem[{Ba{\~n}ados} et~al.(2016){Ba{\~n}ados}, {Venemans}, {Decarli}
  et~al.]{100QSO}
{Ba{\~n}ados} E., {Venemans} B.~P., {Decarli} R., et~al., 2016, \apjs, 227, 11

\bibitem[{Bruns} et~al.(2012){Bruns}, {Wyithe}, {Bland-Hawthorn} \&
  {Dijkstra}]{2012MNRAS.421.2543B}
{Bruns} L.~R., {Wyithe} J.~S.~B., {Bland-Hawthorn} J., {Dijkstra} M., 2012,
  \mnras, 421, 2543

\bibitem[{De Rosa} et~al.(2011){De Rosa}, {Decarli}, {Walter}
  et~al.]{2011ApJ...739...56D}
{De Rosa} G., {Decarli} R., {Walter} F., et~al., 2011, \apj, 739, 56

\bibitem[{Djorgovski} et~al.(2003){Djorgovski}, {Stern}, {Mahabal} \&
  {Brunner}]{2003ApJ...596...67D}
{Djorgovski} S.~G., {Stern} D., {Mahabal} A.~A., {Brunner} R., 2003, \apj, 596,
  67

\bibitem[{Francis} \& {Bland-Hawthorn}(2004)]{2004MNRAS.353..301F}
{Francis} P.~J., {Bland-Hawthorn} J., 2004, \mnras, 353, 301

\bibitem[{Garel} et~al.(2015){Garel}, {Blaizot}, {Guiderdoni}, {Michel-Dansac},
  {Hayes} \& {Verhamme}]{2015MNRAS.450.1279G}
{Garel} T., {Blaizot} J., {Guiderdoni} B., {Michel-Dansac} L., {Hayes} M.,
  {Verhamme} A., 2015, \mnras, 450, 1279

\bibitem[{Gawiser} et~al.(2007){Gawiser}, {Francke}, {Lai}
  et~al.]{2007ApJ...671..278G}
{Gawiser} E., {Francke} H., {Lai} K., et~al., 2007, \apj, 671, 278

\bibitem[{Goto} et~al.(2015){Goto}, {Oi}, {Ohyama} et~al.]{Goto15CFHT}
{Goto} T., {Oi} N., {Ohyama} Y., et~al., 2015, \mnras, 452, 1684

\bibitem[{Goto} \& {Toft}(2015)]{Goto_CO}
{Goto} T., {Toft} S., 2015, \aap, 579, A17

\bibitem[{Goto} et~al.(2009){Goto}, {Utsumi}, {Furusawa}, {Miyazaki} \&
  {Komiyama}]{Goto09}
{Goto} T., {Utsumi} Y., {Furusawa} H., {Miyazaki} S., {Komiyama} Y., 2009,
  \mnras, 400, 843

\bibitem[{Goto} et~al.(2011){Goto}, {Utsumi}, {Hattori}, {Miyazaki} \&
  {Yamauchi}]{Goto11}
{Goto} T., {Utsumi} Y., {Hattori} T., {Miyazaki} S., {Yamauchi} C., 2011,
  \mnras, 415, L1

\bibitem[{Husband} et~al.(2015){Husband}, {Bremer}, {Stanway} \&
  {Lehnert}]{2015MNRAS.452.2388H}
{Husband} K., {Bremer} M.~N., {Stanway} E.~R., {Lehnert} M.~D., 2015, \mnras,
  452, 2388

\bibitem[{Kashikawa} et~al.(2007){Kashikawa}, {Kitayama}, {Doi}, {Misawa},
  {Komiyama} \& {Ota}]{K07}
{Kashikawa} N., {Kitayama} T., {Doi} M., {Misawa} T., {Komiyama} Y., {Ota} K.,
  2007, \apj, 663, 765

\bibitem[{Kikuta} et~al.(2017){Kikuta}, {Imanishi}, {Matsuoka}, {Matsuda},
  {Shimasaku} \& {Nakata}]{kikuta}
{Kikuta} S., {Imanishi} M., {Matsuoka} Y., {Matsuda} Y., {Shimasaku} K.,
  {Nakata} F., 2017, Accepted for publication in ApJ

\bibitem[{Kim} et~al.(2009){Kim}, {Stiavelli}, {Trenti}
  et~al.]{2009ApJ...695..809K}
{Kim} S., {Stiavelli} M., {Trenti} M., et~al., 2009, \apj, 695, 809

\bibitem[{Matthee} et~al.(2015){Matthee}, {Sobral}, {Santos}, {R{\"o}ttgering},
  {Darvish} \& {Mobasher}]{2015MNRAS.451..400M}
{Matthee} J., {Sobral} D., {Santos} S., {R{\"o}ttgering} H., {Darvish} B.,
  {Mobasher} B., 2015, \mnras, 451, 400

\bibitem[{Mazzucchelli} et~al.(2017){Mazzucchelli}, {Ba{\~n}ados}, {Decarli}
  et~al.]{2017ApJ...834...83M}
{Mazzucchelli} C., {Ba{\~n}ados} E., {Decarli} R., et~al., 2017, \apj, 834, 83

\bibitem[{Ouchi} et~al.(2010){Ouchi}, {Shimasaku}, {Furusawa} et~al.]{O10}
{Ouchi} M., {Shimasaku} K., {Furusawa} H., et~al., 2010, \apj, 723, 869

\bibitem[{Overzier}(2016)]{2016A&ARv..24...14O}
{Overzier} R.~A., 2016, Astronomy and Astrophysics Reviews, 24, 14

\bibitem[{Overzier} et~al.(2006){Overzier}, {Bouwens}, {Illingworth} \&
  {Franx}]{2006ApJ...648L...5O}
{Overzier} R.~A., {Bouwens} R.~J., {Illingworth} G.~D., {Franx} M., 2006,
  \apjl, 648, L5

\bibitem[{Overzier} et~al.(2009){Overzier}, {Heckman}, {Tremonti} et~al.]{RO09}
{Overzier} R.~A., {Heckman} T.~M., {Tremonti} C., et~al., 2009, \apj, 706, 203

\bibitem[{Pentericci} et~al.(2011){Pentericci}, {Fontana}, {Vanzella}
  et~al.]{2011ApJ...743..132P}
{Pentericci} L., {Fontana} A., {Vanzella} E., et~al., 2011, \apj, 743, 132

\bibitem[{Schenker} et~al.(2014){Schenker}, {Ellis}, {Konidaris} \&
  {Stark}]{Schenker14}
{Schenker} M.~A., {Ellis} R.~S., {Konidaris} N.~P., {Stark} D.~P., 2014, \apj,
  795, 20

\bibitem[{Schlegel} et~al.(1998){Schlegel}, {Finkbeiner} \& {Davis}]{SFD98}
{Schlegel} D.~J., {Finkbeiner} D.~P., {Davis} M., 1998, \apj, 500, 525

\bibitem[{Shimasaku} et~al.(2005){Shimasaku}, {Ouchi}, {Furusawa}, {Yoshida},
  {Kashikawa} \& {Okamura}]{shimasaku05}
{Shimasaku} K., {Ouchi} M., {Furusawa} H., {Yoshida} M., {Kashikawa} N.,
  {Okamura} S., 2005, \pasj, 57, 447

\bibitem[{Springel} et~al.(2005){Springel}, {White}, {Jenkins}
  et~al.]{2005Natur.435..629S}
{Springel} V., {White} S.~D.~M., {Jenkins} A., et~al., 2005, \nat, 435, 629

\bibitem[{Stark} et~al.(2011){Stark}, {Ellis} \& {Ouchi}]{2011ApJ...728L...2S}
{Stark} D.~P., {Ellis} R.~S., {Ouchi} M., 2011, \apjl, 728, L2

\bibitem[{Stiavelli} et~al.(2005){Stiavelli}, {Djorgovski}, {Pavlovsky}
  et~al.]{2005ApJ...622L...1S}
{Stiavelli} M., {Djorgovski} S.~G., {Pavlovsky} C., et~al., 2005, \apjl, 622,
  L1

\bibitem[{Swinbank} et~al.(2012){Swinbank}, {Baker}, {Barr}, {Hook} \&
  {Bland-Hawthorn}]{2012MNRAS.422.2980S}
{Swinbank} J., {Baker} J., {Barr} J., {Hook} I., {Bland-Hawthorn} J., 2012,
  \mnras, 422, 2980

\bibitem[{Tang} et~al.(2017){Tang}, {Goto}, {Ohyama} et~al.]{TG}
{Tang} J.-J., {Goto} T., {Ohyama} Y., et~al., 2017, accepted for MNRAS

\bibitem[{Taniguchi} et~al.(2005){Taniguchi}, {Ajiki}, {Nagao} et~al.]{T05}
{Taniguchi} Y., {Ajiki} M., {Nagao} T., et~al., 2005, \pasj, 57, 165

\bibitem[{Utsumi} et~al.(2010){Utsumi}, {Goto}, {Kashikawa} et~al.]{U10}
{Utsumi} Y., {Goto} T., {Kashikawa} N., et~al., 2010, \apj, 721, 1680

\bibitem[{Willott} et~al.(2010){Willott}, {Albert}, {Arzoumanian}
  et~al.]{2010AJ....140..546W}
{Willott} C.~J., {Albert} L., {Arzoumanian} D., et~al., 2010, \aj, 140, 546

\bibitem[{Willott} et~al.(2007){Willott}, {Delorme}, {Omont}
  et~al.]{2007AJ....134.2435W}
{Willott} C.~J., {Delorme} P., {Omont} A., et~al., 2007, \aj, 134, 2435

\bibitem[{Willott} et~al.(2013){Willott}, {Omont} \&
  {Bergeron}]{2013ApJ...770...13W}
{Willott} C.~J., {Omont} A., {Bergeron} J., 2013, \apj, 770, 13

\bibitem[{Zheng} et~al.(2006){Zheng}, {Overzier}, {Bouwens}
  et~al.]{2006ApJ...640..574Z}
{Zheng} W., {Overzier} R.~A., {Bouwens} R.~J., et~al., 2006, \apj, 640, 574

\end{thebibliography}
\bibliographystyle{mnras} 




\label{lastpage}

\end{document}